\documentclass[table,xcdraw]{midl} % Include author names
\usepackage{mwe} % to get dummy images
\usepackage{hyperref}
\usepackage[utf8]{inputenc}
\usepackage{slashbox}

\usepackage[font=small,skip=0pt]{caption}
\usepackage{varwidth}
\DeclareCaptionFormat{varwidth}{%
  \begin{varwidth}{\linewidth}#1#2#3\end{varwidth}%
}
\newcommand{\squeezeup}{\vspace{-5.mm}}

\jmlrproceedings{MIDL}{Medical Imaging with Deep Learning}
\jmlrpages{}
\jmlryear{2019}
\jmlrworkshop{MIDL 2019 -- Extended Abstract Track}

\editors{Under Review for MIDL 2019}

\title[SNN, 3D-CNN \& Multi-task loss for TB]{A Multi-Task Self-Normalizing 3D-CNN to Infer Tuberculosis Radiological Manifestations}

\midlauthor{\Name{Pedro M. Gordaliza\nametag{$^{1}$}} \Email{pmacias@ing.uc3m.es}\\
\addr
\Name{Juan José Vaquero\nametag{$^{1}$}} \Email{juanjose.vaquero@uc3m.es}\\
\Name{Sally Sharpe\nametag{$^{2}$}} \Email{sally.sharpe@hpa.org.uk}\\
\addr
\Name{Fergus Gleeson\nametag{$^{3}$}} \Email{fergus.gleeson@oncology.ox.ac.uk}\\
\addr
\Name{Arrate Muñoz-Barrutia\nametag{$^{1}$}} \Email{mamunozb@ing.uc3m.es}
\begin{center}
\addr $^{1}$ Dpto.Bioingenieria e Ing. Aeroespacial, Universidad Carlos III de Madrid, Leganés, Spain \\
\addr $^{2}$ Microbiology Services Division, Public Health England,  Porton Down, England \\
\addr $^{3}$ Oncology Department, The Churchill Hospital, Headington, Oxford, UK 
\end{center}
}

\begin{document}
\squeezeup
\maketitle
\squeezeup
\begin{abstract}
We propose a learning method well-suited to infer the presence of Tuberculosis (TB) manifestations on Computer Tomography (CT) scans mimicking the radiologist reports. Latent features are extracted from the CT volumes employing the \textit{V-Net} encoder and those are the input to a  \textit{Feed-Forward Neural Network (FNN)} for multi-class classification. To overtake the issues (e.g., exploding/vanishing gradients, lack of sensibility) that normally appear when training deep 3D models with datasets of limited size and composed of large volumes, our proposal employs: 1) At the network architecture level, the \textit{scaled exponential linear unit} (SELU) activation which allows the network self-normalization, and 2) at the learning phase, multi-task learning with a loss function weighted by the task \textit{homoscedastic} uncertainty. The results achieve $F_1$-scores close to or above $0.9$ for the detection of TB lesions  and a Root Mean Square Error of $1.16$ for the number of nodules. %Our positive results reflects the promise of our proposal.
\end{abstract}


\section{Introduction}
Tuberculosis (TB) is an infectious disease which generally affects the lungs and has a high incidence and mortality \cite{WorldHealthOrganizationandothers2018Global2018}. Due to the severity of the pandemic, the World Health Organization (WHO) has launched an ambitious plan to eradicate TB by 2030, for which, the extraction of sensitive radiological biomarkers \cite{Nachiappan2017PulmonaryManagement} is a clear need. Traditionally, radiologists through visual inspection of x-ray Computed Tomography (CT) volumes generate reports summarizing the presence of TB-related manifestations. However, using this approach for the extraction of robust TB radiological biomarkers, it is unfeasible and automation required. In recent years, the introduction of deep learning techniques has drastically contributed to this task \cite{Wang2017ChestX-ray8:Diseases, Litjens2017AAnalysis,Hinton2018DeepCare}. For deep learning, knowledge is usually injected into the model in the form of manually segmented masks of the lung lesions. Instead, our proposal directly employs the expertise acquired by the radiologist through years of clinical practice as synthesized in tabular reports. 

%Our methodology is based in a 3D-CNN \cite{Milletari2016V-Net:Segmentation} modified to act as a \textit{Self-Normalizing Neural Network} (SNNs) \cite{Klambauer2017Self-NormalizingNetworks} the extracts distinctive features from whole CT volumes to detect the presence or the quantity of specific TB manifestations. Our deep network is configured as a multi-task model to perform classification acknowledging the fact that TB manifestations do not appear isolated. Uncertainty is used to weight the influence of each task loss \cite{Kendall2018Multi-TaskSemantics}.    

\section{Material and Methods}
%\subsection{Material}
%The \hyperref[sec:expAndResults]{experiments} were accomplished on 
Chest CT scans ($56$) acquired from $14$ male Cynomolgus macaques at $3$, $7$, $11$ and $16$ weeks after TB aerosol exposure were employed. First, the CT volumes were cropped \cite{Gordaliza2018UnsupervisedModelb} and resampled to $1\mbox{mm}\times 1\mbox{mm}\times 2\mbox{mm}$. During the training, data augmentation is performed (elastic transformation, addition of Gaussian noise). As labelled data, we employ the tabular reports elaborated by a radiologist ($20$ years experienced) that contain the number of detected nodules ($0-15$) and binary annotations indicating the presence per lung lobe of the most common TB manifestations (e.g., cavitations, conglomerations, consolidations and trees in bud) \cite{Nachiappan2017PulmonaryManagement}.

\subsection{Model Architecture}\label{ssec:ModelArq}
Our implementation (\figureref{fig:architecture}) extracts latent features from the volumes employing the \textit{V-Net} \cite{Milletari2016V-Net:Segmentation} encoder. The extracted features are used as input of the \textit{Feed-Forward Neural Networks} (FNNs) for multi-class classification. The encoder generates $1,376.256$ features, which feed FNN$_{1}$ (task-shared).
The outputs of FNN$_{1}$ are employed as the input of two independent FNNs, FNN$_R$ and FNN$_B$, corresponding with regression (nodule counting) and binary tasks.
Dropout or Batch Normalization (BN) \cite{Ioffe2015BatchShift} layers are included where is needed. When employing SNN, BN is unnecessary.

\begin{figure}[h]
 % Caption and label go in the first argument and the figure contents
 % go in the second argument
\floatconts
  {fig:architecture}
  {\caption{3D-CNN $+$ three FNNs (\textit{Feed Forward neural networks}): $FNN_1$ (tasks-shared parameters) and, $FNN_R$ and $FNN_B$ for prediction of regression and binary tasks. BN is not present with SELU.}}
  {\includegraphics[width=0.9\linewidth,height=0.8\textheight,keepaspectratio]{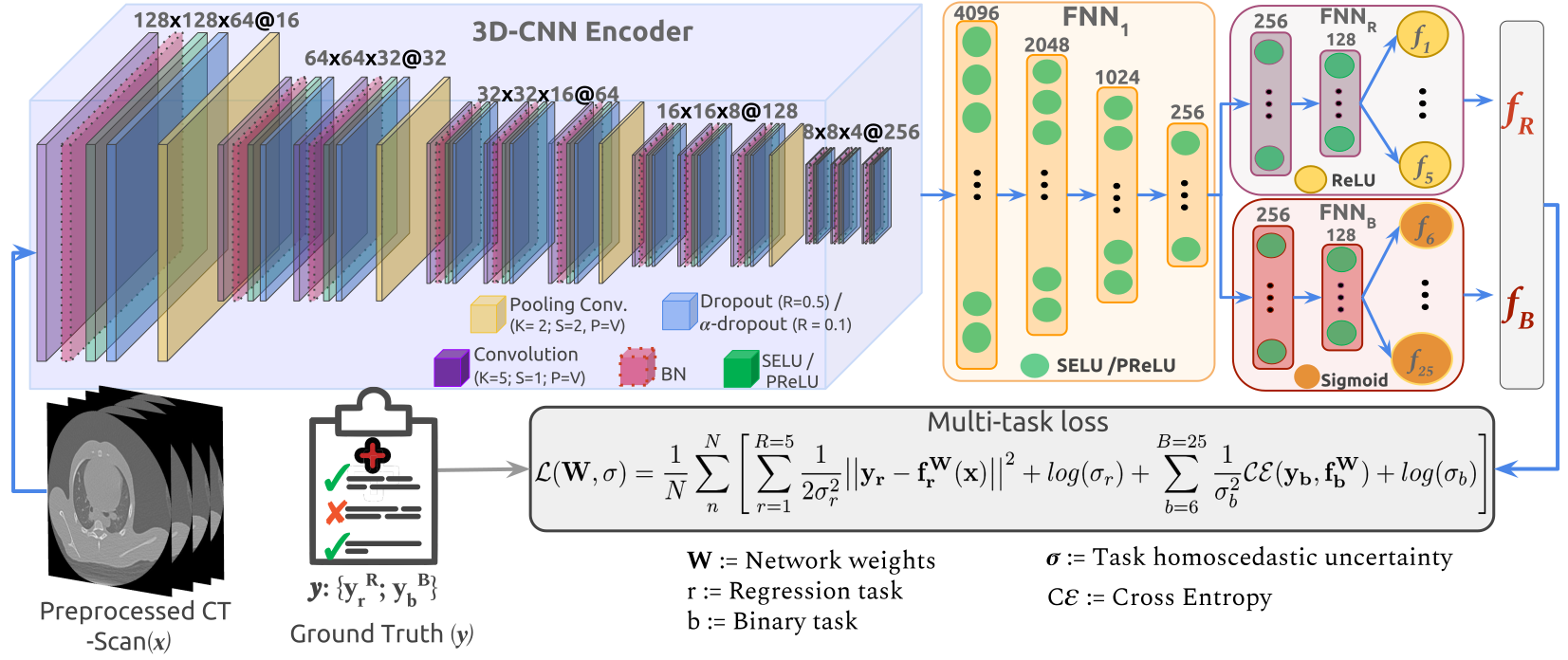}}
\end{figure}
\squeezeup

\subsection{Self-Normalizing Neural Networks}\label{ssec:snn}
 For regularization and normalization purposes, we apply the \textit{Self-Normalizing Neural Networks} (SNNs) strategy to our model \cite{Klambauer2017Self-NormalizingNetworks} that preserves the activation magnitude close to zero mean and unit variance. As activation function, we use the \textit{Scaled Exponential Linear Unit} (SELU) and as dropout, the $\alpha$-dropout \cite{Klambauer2017Self-NormalizingNetworks}.

\subsection{Learning Principle: Uncertainty Weighted Multi-task Loss }\label{ssec:Multi-loss}
When working in multi-task classification, the decision on the influence of each task in the final loss is non-trivial. Traditionally, the loss is computed as  $\mathbf{\mathcal{L}} = \sum_i w_i \mathcal{L}_i$, where the weights $w_i$ are selected either manually or after an exhaustive grid search. This approach is highly influenced by the units and the scale of each task and extremely time-consuming. Recently, \cite{Kendall2018Multi-TaskSemantics} \textit{et al.} proposed a method to compute the weights guiding each specific task loss by \textit{homoscedastic} uncertainty of the predictions.  In this work, we adapt this approach to our multi-label classification problem and derive  a  loss  function  for  our  model (\figureref{fig:architecture}). 

\section{Experiments and Results}\label{sec:expAndResults}
For the evaluation, we employed $5$-fold Cross-Validation (CV): Validation set ($4$ CT volumes of one subject); training set (remaining $13$ subjects). We compared the proposed model (\textit{SELU}) and a modified version which employs a Parametric Rectified Linear Unit (PReLU), Batch Normalization and standard dropout (\textit{BN+PReLU}). The models are trained with $10.000$ iterations of the ADAM optimizer \cite{Kingma2014Adam:Optimization} (learning rate$=10^{-5}$, mini-batch size$=15$). For \textit{BN+PReLU}, standard parameters were employed for the ADAM optimizer and a dropout rate of $0.5$. For \textit{SELU}: $\beta_2=0.9$, $\epsilon=0.01$ and alpha-dropout$=0.1$ \cite{Klambauer2017Self-NormalizingNetworks}. 
\figureref{fig:loss_chart} shows for the validation data, the loss at each fold for the BN+PReLU (in red) and the proposed model (in blue). It can be observed that SELU presents improved convergence. 
The inference error is estimated by the Root Mean Square Error (RMSE) for the nodules count tasks and the $F_{1}$-score for the twenty binary tasks. \tableref{tab:Fold_results} presents the results per fold. No significant  statistical differences were found for paired t-test ($p \nleq 0.05$). Nevertheless, the proposed model presents better \textit{RMSE} and $F_{1}$-score results. 
%In this context, the table present a \textit{RMSE} around the unit ($1.09$-$0.45$) for the nodule counting task and a suitable  $F_{1}$-score binary tasks ($0.98$-$0.79$).

\begin{table}[h]
\centering\rowcolors{2}{gray!6}{white}
\floatconts
  {tab:Fold_results}%
  {\caption{Performance measure results the for BN+PReLU and the proposed model (SELU).}}%

\squeezeup
\resizebox{\linewidth}{!}{
\begin{tabular}{ccccccccccc}
\rowcolor[HTML]{9B9B9B} 
\cellcolor[HTML]{343434}{\color[HTML]{FFFFFF} \textbf{Manifestation/}} & \multicolumn{2}{c}{\cellcolor[HTML]{9B9B9B}\textbf{Nodules} [RMSE]} & \multicolumn{2}{c}{\cellcolor[HTML]{9B9B9B}{\color[HTML]{343434} \textbf{Cavitations [$F_1$]}}} & \multicolumn{2}{c}{\cellcolor[HTML]{9B9B9B}{\color[HTML]{343434} \textbf{Conglomeration} [$F_1$]}} & \multicolumn{2}{c}{\cellcolor[HTML]{9B9B9B}{\color[HTML]{343434} \textbf{Consolidation} [$F_1$]}} & \multicolumn{2}{c}{\cellcolor[HTML]{9B9B9B}\textbf{Tree in bud} [$F_1$]} \\
\cellcolor[HTML]{343434}{\color[HTML]{FFFFFF} \textbf{Fold}} & \cellcolor[HTML]{C0C0C0}BN+PReLU & \cellcolor[HTML]{EFEFEF}SELU & \cellcolor[HTML]{C0C0C0}BN+PReLU & \cellcolor[HTML]{EFEFEF}SELU & \cellcolor[HTML]{C0C0C0}BN+PReLU & \cellcolor[HTML]{EFEFEF}SELU & \cellcolor[HTML]{C0C0C0}BN+PReLU & \cellcolor[HTML]{EFEFEF}SELU & \cellcolor[HTML]{C0C0C0}BN+PReLU & \cellcolor[HTML]{EFEFEF}SELU \\
\multicolumn{1}{c|}{\cellcolor[HTML]{9B9B9B}{\color[HTML]{343434} \textbf{1}}} & $0.73_{0.84}$ & \multicolumn{1}{c|}{$0.85_{0.35}$} & $0.88_{0.11}$ & \multicolumn{1}{c|}{$0.88_{0.12}$} & $0.90_{0.13}$ & \multicolumn{1}{c|}{$0.92_{0.12}$} & $0.88_{0.18}$ & \multicolumn{1}{c|}{$0.83_{0.22}$} & $0.83_{0.22}$ & $0.79_{0.23}$ \\
\rowcolor[HTML]{EFEFEF} 
\multicolumn{1}{c|}{\cellcolor[HTML]{9B9B9B}{\color[HTML]{343434} \textbf{2}}} & $1.15_{0.89}$ & \multicolumn{1}{c|}{\cellcolor[HTML]{EFEFEF}$1.09_{0.83}$} & $0.86_{0.23}$ & \multicolumn{1}{c|}{\cellcolor[HTML]{EFEFEF}$0.88_{0.22}$} & $0.94_{0.17}$ & \multicolumn{1}{c|}{\cellcolor[HTML]{EFEFEF}$0.93_{0.18}$} & $0.93_{0.15}$ & \multicolumn{1}{c|}{\cellcolor[HTML]{EFEFEF}$0.93_{0.18}$} & $0.97_{0.08}$ & $0.97_{0.08}$ \\
\multicolumn{1}{c|}{\cellcolor[HTML]{9B9B9B}{\color[HTML]{343434} \textbf{3}}} & $0.41_{0.34}$ & \multicolumn{1}{c|}{$0.23_{0.39}$} & $0.85_{0.12}$ & \multicolumn{1}{c|}{$0.87_{0.11}$} & $0.89_{0.19}$ & \multicolumn{1}{c|}{$0.97_{0.11}$} & $0.96_{0.11}$ & \multicolumn{1}{c|}{$0.98_{0.03}$} & $0.95_{0.14}$ & $0.96_{0.12}$ \\
\rowcolor[HTML]{EFEFEF} 
\multicolumn{1}{c|}{\cellcolor[HTML]{9B9B9B}{\color[HTML]{343434} \textbf{4}}} & $1.22_{0.6}$ & \multicolumn{1}{c|}{\cellcolor[HTML]{EFEFEF}$0.78_{0.74}$} & $0.94_{0.15}$ & \multicolumn{1}{c|}{\cellcolor[HTML]{EFEFEF}$0.9_{0.19}$} & $0.9_{0.19}$ & \multicolumn{1}{c|}{\cellcolor[HTML]{EFEFEF}$0.92_{0.15}$} & $0.88_{0.18}$ & \multicolumn{1}{c|}{\cellcolor[HTML]{EFEFEF}$0.87_{0.18}$} & $0.87_{0.21}$ & $0.94_{0.15}$ \\
\multicolumn{1}{c|}{\cellcolor[HTML]{9B9B9B}{\color[HTML]{343434} \textbf{5}}} & $0.41_{0.8}$ & \multicolumn{1}{c|}{$0.45_{0.8}$} & $0.93_{0.17}$ & \multicolumn{1}{c|}{$0.94_{0.17}$} & $0.94_{0.14}$ & \multicolumn{1}{c|}{$0.96_{0.12}$} & $0.90_{0.18}$ & \multicolumn{1}{c|}{$0.94_{0.14}$} & $0.91_{0.18}$ & $0.92_{0.17}$ \\ \hline
\rowcolor[HTML]{EFEFEF} 
\multicolumn{1}{c|}{\cellcolor[HTML]{9B9B9B}{\color[HTML]{343434} \textbf{Total}}} & $0.78_{0.69}$ & \multicolumn{1}{c|}{\cellcolor[HTML]{EFEFEF}$\mathbf{0.68_{0.62}}$} & $0.89_{0.16}$ & \multicolumn{1}{c|}{\cellcolor[HTML]{EFEFEF}$\mathbf{0.90_{0.17}}$} & $0.91_{0.16}$ & \multicolumn{1}{c|}{\cellcolor[HTML]{EFEFEF}$\mathbf{0.94_{0.16}}$} & $0.91_{0.16}$ & \multicolumn{1}{c|}{\cellcolor[HTML]{EFEFEF}$0.91_{0.15}$} & $0.91_{0.17}$ & $\mathbf{0.92_{0.15}}$
\end{tabular}
}
\end{table}
\squeezeup

\begin{figure}[h]
 % Caption and label go in the first argument and the figure contents
 % go in the second argument
\floatconts
  {fig:loss_chart}
  {\caption{Evolution of the validation loss for each fold of the Cross-Validation.}}
  {\includegraphics[width=0.8\linewidth,height=\textheight,keepaspectratio]{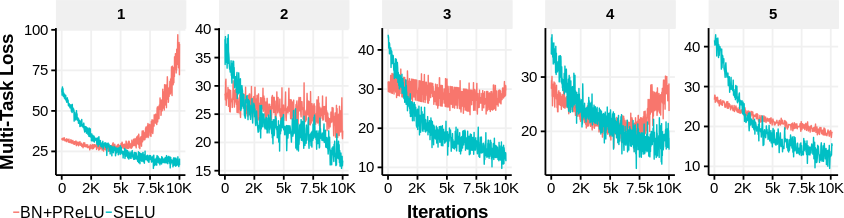}}
\end{figure}

\section{Conclusions}
 Although further validation on large datasets is needed, the work presents a promising inference of the radiologist reports. This is achieved with a reduced computational complexity by avoiding normalization layers and hyperparameter tuning of the loss weights. %The results are consistent with known facts about the disease. The inference of the severely diseased subjects' reports are poorer when compared to those moderately affected by the disease, as can be observed for subject $\#3$ and  $\#2$.
%We acknowledge that the employed dataset is limited, and further validation is needed. %Nevertheless, our methodology for multi-label classification, enables the inclusion of \textit{SNNs} within 3D CNNs and multi-task learning guided by the uncertainty.

\newpage
% Acknowledgments---Will not appear in anonymized version
\midlacknowledgments{The research leading to these results received funding from the Innovative Medicines Initiative (www.imi.europa.eu) Joint Undertaking under grant agreement no. 115337, whose resources comprise funding from EU FP7/2007-2013 and EFPIA companies in kind contribution. This work was partially funded by projects RTC-2015-3772-1, TEC2015-73064-EXP and TEC2016-78052-R from the Spanish Ministry of Economy, Industry and Competitiveness (MEIC), TOPUS S2013/MIT-3024 project from the regional government of Madrid and by the Department of Health, UK.\\
This material is based upon work supported by Google Cloud Platform.
}

\bibliography{mendeley.bib}
\end{document}